# Nonlocal effects and enhanced nonreciprocity in current-driven graphene systems

Tiago A. Morgado[1], Mário G. Silveirinha[1,2*]

[1]*Instituto de Telecomunicações and Department of Electrical Engineering, University of Coimbra, 3030-290 Coimbra, Portugal*

[2]*University of Lisbon, Instituto Superior Técnico, Avenida Rovisco Pais, 1, 1049-001 Lisboa, Portugal*

E-mail: tiago.morgado@co.it.pt, mario.silveirinha@co.it.pt

**Abstract**

A graphene sheet biased with a drift electric current offers a tantalizing opportunity to attain unidirectional, backscattering-immune, and subwavelength light propagation, as proposed in [T. A. Morgado, M. G. Silveirinha, *ACS Photonics* **5**(*11*), 4253 (2018)]. Here, we investigate in detail the impact of the intrinsic nonlocal response of graphene in the dispersion characteristics of the current-driven plasmons supported by single-layer and double-layer graphene systems. It is theoretically shown that even though the nonlocal effects weaken the spectral asymmetry of the plasmons dispersion, the studied platforms can support unidirectional backscattering-immune guided modes. Our analysis also confirms that the drift-current bias can effectively pump the graphene plasmons and enhance the propagation distance. Moreover, it is shown that the nonreciprocity and optical isolation can be boosted by pairing two drift-current biased graphene sheets due to the enhanced radiation drag by the drifting electrons.

---

[*] To whom correspondence should be addressed: E-mail: mario.silveirinha@co.it.pt

# I. INTRODUCTION

Reciprocity is a fundamental feature of conventional photonic systems stemming from the linearity and invariance of Maxwell's equations under time reversal symmetry [1-4]. Lorentz reciprocity implies that the transmission level between two points is the same independent of the propagation direction [2-3].

Nonreciprocal devices are, however, essential building blocks of many photonic systems. Typical nonreciprocal photonic devices rely on magneto-optical materials externally biased by a static magnetic field [5]. Curiously, some of these systems have nontrivial topological properties and support unidirectional backscattering-immune chiral edge states [6-11]. As widely discussed in the recent literature, nonreciprocal "magnetic" solutions are not fully satisfactory, as the associated biasing circuit is bulky and the nonreciprocal responses of magneto-optical materials are relatively weak at terahertz and optical frequencies. Due to these reasons, there has been an increasing demand for magnetic-free nonreciprocal photonic components that can be straightforwardly incorporated in highly-integrated photonic circuits [12-24].

Recently, a novel route to achieve magnetic-free nonreciprocal subwavelength propagation in graphene that is compatible with modern highly-integrated nanophotonic technology was explored in Refs. [25-30]. In particular, we theoretically demonstrated in [29] that a graphene sheet biased with a drift electric current may enable the unidirectional broadband propagation of surface plasmon polaritons (SPPs). Furthermore, we have also shown that the current-driven graphene plasmons are protected against backscattering from obstacles and imperfections, similar to the "one-way" topologically protected chiral edge modes supported by nonreciprocal topological photonic systems [6-11].

The impact of the bare nonlocal graphene response, i.e., the dependence of the conductivity on the wave vector, in the dispersion of the current-driven graphene plasmons was only superficially analyzed in Ref. [29]. However, nonlocal effects may critically affect the propagation of the graphene plasmons, especially for highly confined SPPs with large wavevectors approaching the Fermi wavenumber $k_F = \mu_c/(\hbar v_F)$ ($\mu_c$ is the chemical potential, $v_F$ the Fermi velocity, and $\hbar$ is the reduced Planck constant) [31]. In this article, we study in detail the impact of the nonlocal effects in the dispersion characteristics of the current-driven SPPs not only in single-layer graphene (SLG) systems, but also in double-layer graphene (DLG) systems. The motivation to consider a DLG configuration is to examine whether the nonreciprocal effects can be enhanced by having an additional drift-current biased graphene sheet. It is shown – with the nonlocal effects fully taken into account – that for sufficiently large drift velocities the drift current can effectively drag the SPPs, leading to a "one-way" propagation with enhanced propagation lengths both in SLG and DLG systems. Moreover, we show that the nonreciprocity and optical isolation in DLG systems may be indeed boosted as compared to the SLG configuration. These results confirm thereby the exciting potentials of the drift-current biased graphene platforms in nanophotonics.

This paper is organized as follows. In Sec. II the nonlocal conductivity of the drift-current biased graphene is reviewed and discussed. In Sec. III, we analyze in detail the influence of the nonlocal effects on the dispersion of the current-driven graphene plasmons supported by a single-layer graphene system. In Sec. IV, we investigate the dispersion properties of the current-driven SPPs in a double-layer graphene system. The conclusions are drawn in Sec. V.

## II. NONLOCAL CONDUCTIVITY OF THE DRIFT-CURRENT BIASED GRAPHENE

The nonlocal response of graphene can be characterized using the random-phase approximation (RPA) [32-34], which yields an analytical expression for the surface conductivity of graphene in the zero-temperature limit [32, 34-35]. For a positive real-valued frequency $\omega$ and a real-valued transverse wave number $q = \sqrt{k_x^2 + k_y^2}$, the collisionless nonlocal graphene conductivity is given by (the space-time variation $e^{i(k_x x + k_y y)} e^{-i\omega t}$ is implicit, with $\omega$ being the oscillation frequency) [34-35]:

$$\sigma_g(\omega, q) = -i\omega \frac{e^2}{4\pi\hbar} \left[ \frac{8k_F}{q^2 v_F} + \frac{\left(G(-\Delta_-) - i\pi\right)\theta(-\Delta_- - 1) + G(\Delta_-)\theta(\Delta_- + 1) - \left(G(\Delta_+) - i\pi\right)}{\sqrt{\omega^2 - q^2 v_F^2}} \right],$$

(1)

with $G(z) = z\sqrt{z^2 - 1} - \ln(z + \sqrt{z^2 - 1})$, $\Delta_\pm = (\hbar\omega \pm 2\mu_c)/(\hbar v_F q)$, and $\theta(x)$ is the unit step function. This formula includes both the intraband and interband contributions. The effect of collisions can be taken into account using the phenomenological formula [31, 34-35]

$$\sigma_g(\omega, q, \tau) = -i\omega \frac{\left(1 + i/(\omega\tau)\right) \chi(\omega + i\tau, q)}{1 + i/(\omega\tau) \chi(\omega + i\tau, q)/\chi(0, q)}, \quad (2)$$

where $\chi(\omega, q) = \sigma_g(\omega, q)/(-i\omega)$ is the susceptibility and $\tau$ is the relaxation time.

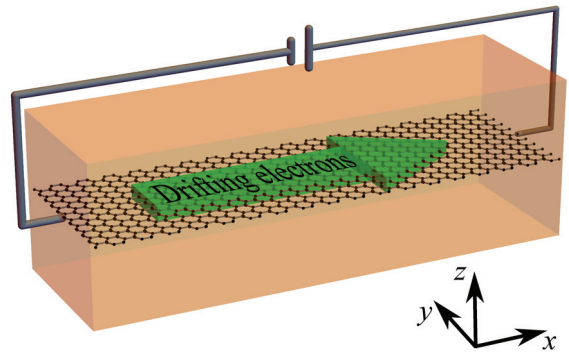

**Fig. 1.** A graphene sheet embedded in a dielectric with permittivity $\varepsilon_{r,s}$. A static voltage generator induces an electron drift in the graphene sheet.

According to the Galilean-Doppler shift model developed in [36-37], when the graphene sheet is biased by a drift current the conductivity should be modified as

$$\sigma_{\mathrm{g}}^{\mathrm{drift}}\left(\omega, k_{x}\right) \approx \left(\omega/\tilde{\omega}\right) \sigma_{\mathrm{g}}\left(\tilde{\omega}, q\right)\Big|_{q=\sqrt{k_x^2}}. \qquad (3)$$

The drifting electrons travel along the $x$-direction with drift velocity $v_0$ [see Fig. 1], and it is implicit that the in-plane electric field is along $x$ (longitudinal excitation). In the above, $\tilde{\omega} = \omega - k_x v_0$ is the Doppler-shifted frequency, $k_x$ is the wave number along $x$-direction (we only consider plasmons with $k_y = 0$), and $\sigma_{\mathrm{g}}(\omega, q)$ is the nonlocal zero-temperature graphene conductivity given by Eqs. (1-2).

Due to the dissipative response of graphene [38], the graphene plasmons are excitations with complex-valued wave number $k_x = k_x' + i k_x''$. In order to numerically calculate $\sigma_{\mathrm{g}}^{\mathrm{drift}}$ it is necessary to evaluate the bare graphene conductivity $\sigma_{\mathrm{g}}$ for $\tilde{\omega}$ and $k_x$ complex-valued. Crucially, this requires that Eq. (1) is surgically modified to become an analytic function of $k_x$. In particular, it is necessary to adjust the branch-cuts of the square root and logarithm functions so that they become continuous for complex-valued $k_x$. In the Appendix A, we provide an explicit formula for $\sigma_{\mathrm{g}}$ valid for moderately small deviations of $k_x$ from the real axis ($|k_x''/k_x'| \ll 1$). In addition, we also show how one can modify Eq. (1) so that it becomes valid for complex frequencies and a real-valued $k_x$.

Throughout the paper, it is assumed that $k_{\mathrm{B}} T \ll \mu_{\mathrm{c}}$, so that the zero-temperature ($T = 0\,\mathrm{K}$) approximation can be used to determine the graphene surface conductivity. The chemical potential of the graphene sheet is taken equal to $\mu_{\mathrm{c}} = 0.1\,\mathrm{eV}$, the collision time $\tau = 170\,\mathrm{fs}$ [28], and it is supposed that the graphene sheet is surrounded by a dielectric with relative permittivity $\varepsilon_{\mathrm{r,s}} = 4$ (e.g., $SiO_2$ or h-BN).

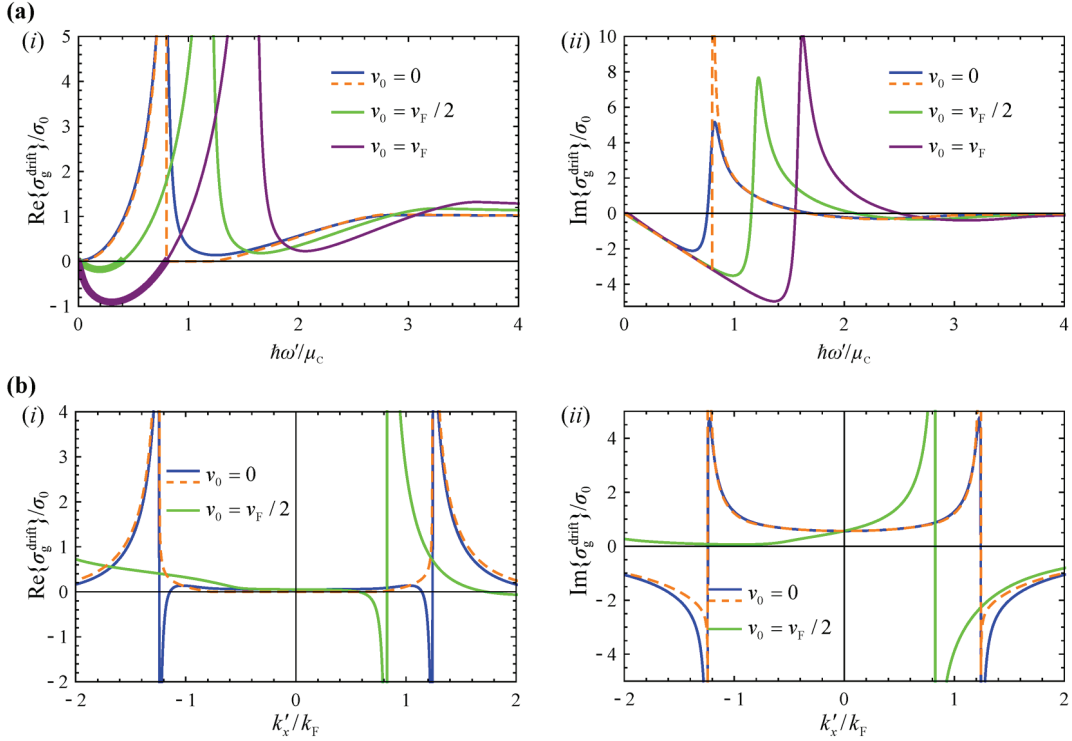

**Fig. 2.** (a) Real (*i*) and imaginary (*ii*) parts of the graphene conductivity in the upper-half frequency-plane as a function of the normalized frequency $\hbar\omega'/\mu_c$ ($\omega = \omega' + i\omega''$) for different drift velocities, $k_x = 0.8 k_F$, $\omega''/(2\pi) = 0.1\,\mathrm{THz}$, and calculated using Eq. (A2). The thicker green and purple curves in (*i*) correspond to the negative Landau damping regions. (b) Real (*i*) and imaginary (*ii*) parts of the graphene conductivity as a function of $k'_x/k_F$ ($k_x = k'_x + i k''_x$) for different drift velocities, $\omega/(2\pi) = 30\,\mathrm{THz}$, $k''_x = 0.05 k'_x$, and calculated using Eq. (A1). The orange dashed curves in (a) and (b) correspond to the collisionless ($\tau \to \infty$) nonlocal graphene conductivity with real-valued $\omega$ and $k_x$, calculated using Eq. (1). The conductivity normalization factor in the plots is $\sigma_0 = e^2/(4\hbar)$.

Figure 2 illustrates the variation of the nonlocal graphene conductivity with either $\omega$ or $k_x$ complex-valued, calculated using Eqs. (A1)-(A2) [solid curves in Fig. 2]. All the curves are analytical continuations of the no-drift collisionless graphene conductivity result [given by Eq. (1)] represented with orange dashed curves in Fig. 2. In general, when $v_0 = 0$ the graphene conductivity exhibits a resonant behavior at $\omega' \approx k_x v_F$ [e.g., see the blue solid and orange dashed curves of Figs. 2(a)(*i*)-(*ii*)], which is the well-known square-root singularity of the graphene conductivity at the threshold of single-particle excitations or Landau damping [32, 39-40]. Crucially, the drift-current biased graphene is an active medium, as one can have $\mathrm{Re}\{\sigma_g^{\mathrm{drift}}\} < 0$ in the upper-half frequency plane ($\omega'' > 0$) for $k_x$ real-valued [36] [see thicker green and purple curves in Fig. 2(a)(*i*)]. This regime, which is fully compatible with the square root singularity of the bare-graphene response, corresponds to a "negative Landau damping". The optical gain is due to the conversion of kinetic energy of the moving charges into electromagnetic radiation.

The drift current ($v_0 \neq 0$) shifts the position of the square-root singularity due to the frequency Doppler shift. Furthermore, the drift current leads to an evident symmetry breaking (nonreciprocity) of the graphene response [see Figs. 2(b)(*i*)-(*ii*)] such that $\sigma_g^{\text{drift}}(\omega, k_x') \neq \sigma_g^{\text{drift}}(\omega, -k_x')$, in accordance with Eq. (3).

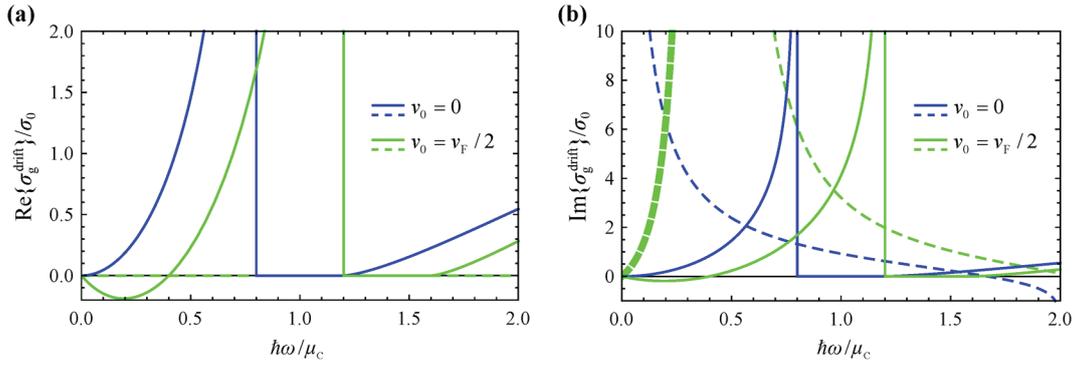

**Fig. 3.** (a) Real (a) and imaginary (b) parts of the collisionless graphene conductivity as a function of the normalized frequency $\hbar\omega/\mu_c$ for $k_x = 0.8 k_F$. The conductivity normalization factor is $\sigma_0 = e^2/(4\hbar)$. Solid curves: nonlocal graphene conductivity (Eq. (1)); Dashed curves: local Kubo conductivity [31, 34]. The thicker dashed green curve in (b) corresponds to the anomalous dispersion region. Note that the real part of the conductivity calculated using the local conductivity is exactly zero for $\omega < 2\mu_c/\hbar$, and thus the dashed lines in panel (a) are coincident with the horizontal axis.

Figure 3 compares the conductivity dispersion obtained with Eq. (3) (solid curves), with the result predicted by a local (long wavelength) approximation $\sigma_g^{\text{drift}}(\omega, k_x) \approx (\omega/\tilde{\omega}) \sigma_g(\tilde{\omega}, q = 0^+)$ (dashed curves). Note that $\sigma_g(\omega, q = 0^+)$ is the (bare) graphene conductivity determined by the standard local Kubo's formula [31, 34]. For simplicity, the effect of collisions is ignored so that $\tau \to \infty$. The conductivity is evaluated for $\omega$ and $k_x$ real-valued. As seen in Fig. 3(a), the local model yields $\text{Re}\{\sigma_g^{\text{drift}}(\omega)\}=0$) for frequencies below the interband transition threshold $\omega=2\mu_c/\hbar$, even in presence of the drift current (dashed curves). In contrast, the nonlocal model predicts the negative Landau damping region with $\text{Re}\{\sigma_g^{\text{drift}}(\omega)\}<0$, as discussed previously. Thereby, the active response of the drift-biased graphene sheet is less evident in the local model formalism. Interestingly, Fig. 3(b) shows that with a drift-current biasing, $\text{Im}\{\sigma_g^{\text{drift}}(\omega)\}$ calculated with the local conductivity [green dashed curve] exhibits an anomalous dispersion, with $\text{Im}\{\partial_\omega \sigma_g^{\text{drift}}(\omega)\}>0$, for low frequencies [see thicker green curve in Fig. 3(b)]. We show in Appendix B, that the anomalous conductivity dispersion implies that the energy "stored" in the material is *negative*. In other words, when the material interacts with a time-harmonic excitation it *gives away* energy before reaching a steady state, rather than extracting energy from the excitation as in the no-drift case. This property unveils the active nature of the drift-biased graphene in the local-model description.

### III. PLASMONS IN A DRIFT-CURRENT BIASED GRAPHENE SHEET

The dispersion characteristic of the transverse magnetic (TM) current-driven graphene SPPs in a single-layer graphene system (Fig. 1) is given by [29-31]

$$2i\omega\varepsilon_0\varepsilon_{r,s}/\gamma_s - \sigma_g^{\text{drift}}(\omega,k_x) = 0, \qquad (4)$$

where $c$ is the speed of light in vacuum, and $\gamma_s = \sqrt{k_x^2 - \varepsilon_{r,s}(\omega/c)^2}$ is the transverse (along $z$) attenuation constant that determines the confinement of the graphene plasmons. Figures 4(a)(*i*)-(*ii*) show the dispersions of the current-driven graphene SPPs calculated using the (bare) nonlocal conductivity ($\sigma_g(\omega,q)$) and the (bare) local conductivity ($\sigma_g(\omega,q=0^+)$) in Eq. (3), respectively. Comparing Fig. 4(a)(*i*) with Fig. 4(a)(*ii*), it is seen that the nonlocal effects may substantially alter the dispersion of the graphene plasmons with values of $k_x$ larger than or comparable to $k_F$. In particular, the nonlocal effects prevent that the SPP dispersion curves cross the line $\omega = 0$ for large drift velocities, different from the local conductivity results. The $\omega = 0$ band crossing predicted by the local conductivity model seems to be rooted on the anomalous dispersion of $\text{Im}\{\sigma_g^{\text{drift}}(\omega)\}$ in the low-frequency regime. Despite this qualitative difference, the drift-current biasing of graphene causes a strong symmetry breaking in the SPPs dispersion in both models. The degree of asymmetry increases as the drift velocity increases. Remarkably, for large enough drift velocities $v_0$, the moving electrons effectively drag the plasmons towards the $+x$ direction (the direction of $v_0$), such that the propagation along the $-x$ direction is forbidden. In the nonlocal model this effect is only possible for drift velocities $v_0 > v_F/2$ [see Fig. 4(a)(*i*)], whereas with a local response it would be feasible for drift velocities as low as $v_0 = v_F/4$ [see Fig. 4(a)(*ii*)].

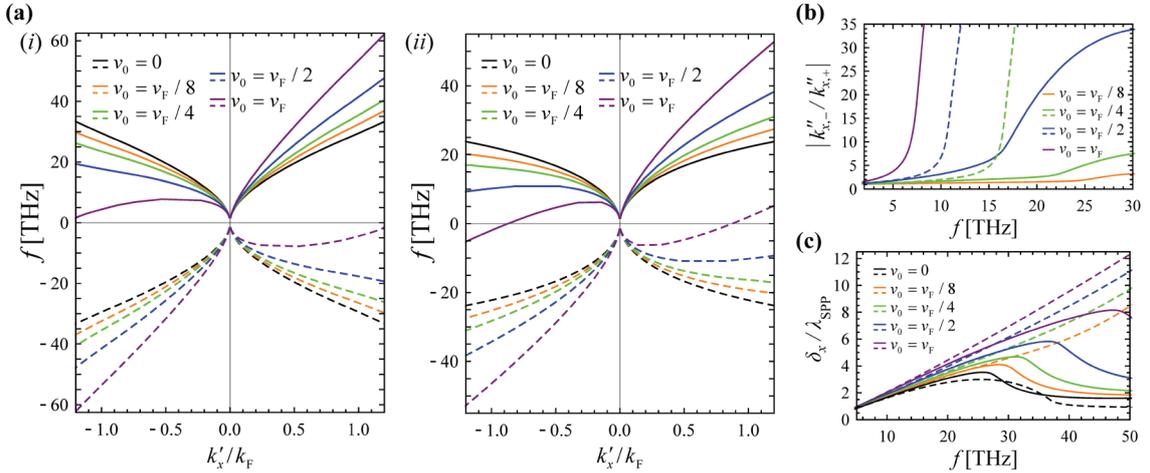

**Fig. 4.** (a) Dispersion of the SPPs supported by the graphene sheet for several drift velocities $v_0$. The dispersion curves are shown for both positive and negative frequencies. (*i*) nonlocal model results calculated using the conductivity formula (A1); (*ii*) local model results calculated using the Kubo conductivity formula [31, 34]. (b) Ratio between the attenuation constants of the SPPs that propagate along the $-x$ and $+x$ directions as a function of the frequency and for several drift velocities $v_0$. Solid curves: nonlocal results; dashed curves: local results. (c) SPP propagation length normalized to the SPP wavelength ($\lambda_{\text{SPP}} = 2\pi/k_x'$) as a function of the frequency for the SPPs propagating along the $+x$

direction (the corresponding SPP dispersions are depicted in (a) and (b)). Solid curves: nonlocal results; dashed curves: local results.

For $v_0 \leq v_F/2$ the SPP dispersion obtained with the nonlocal model predicts both positive and negative $k_x$-solutions for a given frequency [see Fig. 4(a)(*i*)]. Importantly, it turns out that the SPPs that propagate along the $-x$ direction are much more attenuated than the SPPs that propagate along the $+x$ direction, even when the nonlocal effects are considered [see Fig. 4(b)]. Thus, regimes of effective unidirectional propagation may be attainable when $v_0$ is significantly lower than $v_F$, as discussed in greater detail below.

To further characterize the current-driven graphene plasmons, we calculated the SPP propagation length $\delta_x = 1/\text{Im}\{k_x\}$ [Fig. 4(c)]. Notably, the propagation length of the graphene SPPs may be enhanced by the drift current bias. For instance, for $v_0 = v_F/4$ the distance travelled by the SPP can be twice as large than in the no-drift case [see green and black solid curves of Fig. 4(c)], increasing up to 3-5 times as the drift velocity approaches values in the range of $v_F/2$ and $v_F$ [see blue and purple solid curves of Fig. 4(c)]. The local conductivity model tends to overestimate the propagation length enhancement [see dashed curves of Fig. 4(c)]. The enhancement of the propagation length is rooted in the optical gain provided by the drift-current biased graphene.

So far, our analysis was based on the conductivity model (3), where the effect of the drift current is determined by a Galilean Doppler shift. Different models for the graphene conductivity were introduced by other authors [25-28]. In particular, several works characterized the graphene conductivity using a skewed Fermi distribution in the Lindhard formula, which leads to a conductivity that is approximately determined by a relativistic-type Doppler shift transformation. The corresponding conductivity is given by:

$$\sigma_g^{\text{Rel}}(\omega, k_x) \approx (\omega/\tilde{\omega})\sigma_g(\tilde{\omega}, \tilde{q})\big|_{\tilde{q}=\sqrt{\tilde{k}_x^2}}, \quad (5)$$

with $\tilde{\omega} = \gamma(\omega - k_x v_0)$, $\tilde{k}_x = \gamma(k_x - \omega v_0/v_F^2)$, and $\gamma = 1/\sqrt{1 - v_0^2/v_F^2}$ the graphene Lorentz factor. In a single band approximation, the above conductivity formula captures well the main features of the conductivity models proposed in Refs. [26-28] (for more details see [37] and Appendix C).

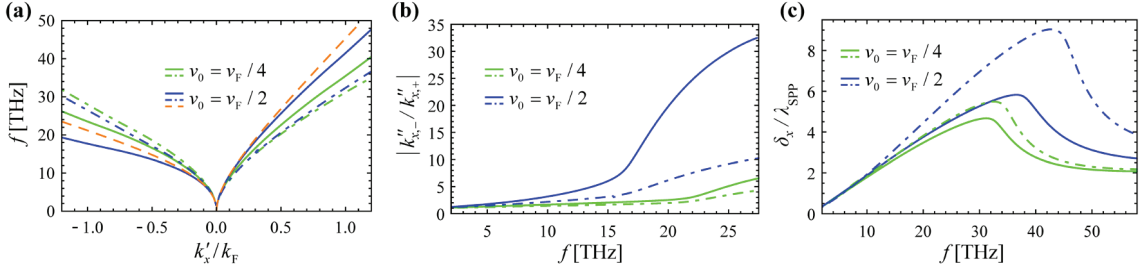

**Fig. 5.** The same as in Fig. 4, but calculated using: solid curves: the Galilean Doppler shift model [Eq. (3)]; dot-dashed curves: the relativistic Doppler shift model [Eq. (5)]. The orange dashed line in (a) was obtained using Eq. (3) with the (bare) graphene conductivity calculated with Eq. [C2], which only takes into account the intraband response.

Figure 5(a) shows the dispersion characteristic of the current-driven SPPs for two different drift velocities, calculated using the Galilean Doppler shift model [Eq. (3)] (solid curves) and the relativistic Doppler shift model [Eq. (5)] (dot-dashed curves). Clearly, the Galilean Doppler shift model predicts a larger spectral asymmetry and stronger nonreciprocity than the relativistic Doppler shift model. Similarly, Fig. 5(b) shows that the relativistic model predicts a weaker discrepancy between the attenuation constants $|k''_{x,-}|$ and $|k''_{x,+}|$ of counter-propagating plasmons. On the other hand, the relativistic-type model predicts a considerably larger enhancement of the propagation length of the SPPs propagating along the direction of the drift-current bias [see Fig. 5(c)]. In our understanding, the Galilean Doppler shift model is more accurate than the relativistic Doppler shift model, as the effect of the drift-current does not change the distribution of canonical momentum (i.e., it does not lead to a skewed a distribution of canonical momentum), but rather the mean energy of the graphene electrons [37].

The nonreciprocal response of graphene arises mainly due to the intraband transitions. This is confirmed by Fig. 5(a), where we plot the SPP dispersion for $v_0 = v_F/2$, calculated using the Galilean Doppler-shift model with the (bare) graphene conductivity given by Eq. [C2] (orange dashed line); this model only takes into account the intraband light-matter interactions. As seen, the calculated dispersion is nearly coincident with the one obtained using the full (intraband + interband) response of graphene (solid blue line in Fig. 5(a)); thus, the interband conductivity term is of secondary importance.

Next, we consider a scenario wherein a linearly polarized emitter placed in the immediate vicinity of the graphene sheet (i.e., in the near-field region) is used to excite the graphene plasmons [see the top panel of Fig. 6(a)]. The SPP fields are calculated with the same formalism as in Ref. [29], but now using the nonlocal conductivity given by formula (A1). Figures 6(a)(*i*)-(*ii*) show time snapshots of the longitudinal component of the electric field ($E_x$) excited by the near-field emitter. The time snapshots are qualitatively similar to those reported in Ref. [29], and confirm that a drift-current biased graphene sheet supports unidirectional SPPs that propagate only along the direction of the drifting electrons ($+x$ direction), with the nonlocal effects fully considered. Moreover, it is interesting to see that one-way propagation regimes are, in fact, possible for drift velocities as low as $v_0 = v_F/4$ [see Fig. 6(a)(*ii*)].

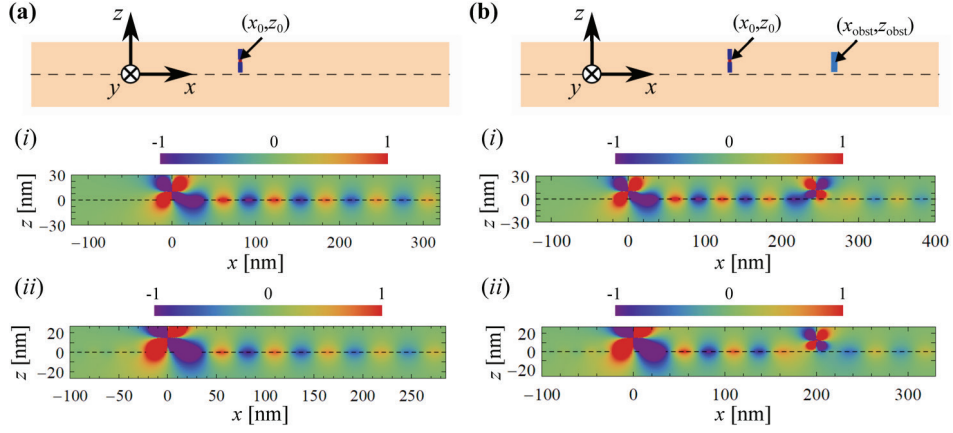

**Fig. 6.** A linearly polarized emitter (with vertical polarization) positioned at $(x,z)=(0,z_0)$ radiates near a graphene sheet when (a) the propagation path is free; (b) an obstacle blocks the propagation path. (*i*)-(*ii*) Time snapshots of the *x*-component of the electric field $E_x$ (in arbitrary unities) of the current-driven graphene SPPs. (*i*) $v_0 = v_F/2$, $f = 31.5$ THz, and $z_0 = 10$ nm; (*ii*) $v_0 = v_F/4$, $f = 29.5$ THz, and $z_0 = 15$ nm. In (b) the obstacle is at (*i*) $(x_{obst}, z_{obst}) = (245\text{ nm}, 12\text{ nm})$, (*ii*) $(x_{obst}, z_{obst}) = (200\text{ nm}, 12\text{ nm})$.

We also studied a configuration in which an obstacle partially blocks the propagation path [see the top panel of Fig. 6(b)]. The obstacle is modeled as a thin metallic strip with $\varepsilon_{r,obst} = -4.1$ and width $w_{obst} = 12$ nm. The electromagnetic fields scattered by the metallic strip are found as explained in Ref. [29]. The time snapshots of the electric field are shown in Fig. 6(b)(*i*)-(*ii*) and confirm that the one-way SPPs are strongly immune to the backscattering effects. Notably, the backscattering suppression is achievable for drift velocities significantly lower than $v_F$ [see Fig. 6(b)(*ii*)]. The obstacle attenuation $\left|E_x^{\text{no-obst}}/E_x^{\text{obst}}\right|$ at 20 THz is shown in Fig. 7 as a function of the drift velocity. The backscattering is evidently reduced for $v_0 > 0.4 v_F$. Qualitatively similar results were reported in Ref. [29] but relying on the local model approximation.

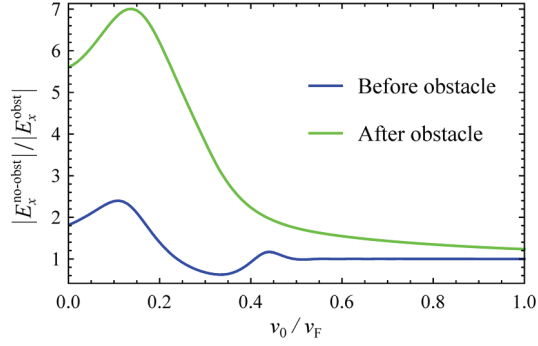

**Fig. 7.** Ratio between the amplitudes of the electric field without the obstacle ( $E_x^{\text{no-obst}}$ ) and with the obstacle ( $E_x^{\text{obst}}$ ) as a function of the normalized drift velocity $v_0/v_F$ at the fixed frequency $f = 20$ THz. The emitter is placed at $(x_0, z_0) = (0, 12 \text{ nm})$ and the obstacle at $(x_{\text{obst}}, z_{\text{obst}}) = (200 \text{ nm}, 12 \text{ nm})$. The field is measured at the point $(x_p, z_p) = (100 \text{ nm}, 0)$ (blue curve) and at the point $(x_p, z_p) = (350 \text{ nm}, 0)$ (green curve).

## IV. PLASMONS IN A DRIFT-CURRENT BIASED DOUBLE-LAYER GRAPHENE SYSTEM

One may wonder if by adding an additional drift-current biased graphene layer to the system it may be possible to enhance the drag of radiation by the drifting electrons. To investigate this problem, next we study the propagation of the current-driven SPPs supported by a double-layer graphene (DLG) system [see Fig. 8(a)]. The dispersion equation for the TM SPPs in this system can be written as [31]

$$\left(\left(2\varepsilon_{\text{r,s}}/\gamma_{\text{s}}\right) - \left(1/\kappa_{\text{g}}^{\text{drift}}\right)\right)^2 e^{\gamma_{\text{s}} d} - \left(1/\kappa_{\text{g}}^{\text{drift}}\right)^2 e^{-\gamma_{\text{s}} d} = 0, \tag{6}$$

where $d$ is the distance between the two graphene sheets, and $\kappa_{\text{g}}^{\text{drift}}(\omega, k_x) = i\omega\varepsilon_0 / \sigma_{\text{g}}^{\text{drift}}(\omega, k_x)$. The two graphene sheets are simultaneously biased with the same drift current and with the same chemical potential ( $\mu_{\text{c}} = 0.1$ eV ).

The calculated dispersion characteristic for a distance $d = 5$ nm and different drift velocities is shown in Fig. 8(b). As expected, two hybridized SPP modes arise in this system: an acoustic (or even) mode and an optical (or odd) mode; the designations even and odd refer to the magnetic field profile. Without the drift current biasing [solid and dashed black curves in Fig. 8(b)(*i*)], the dispersion curves of the two SPP modes are related as $\omega(k'_x) = \omega(-k'_x)$, in agreement with the parity and time-reversal symmetries. Quite differently, in the presence of a drift-current biasing, the parity and time-reversal symmetries are individually broken, and thus the spectral symmetry between the $+k'_x$ and $-k'_x$ parts of the dispersion curves is lost [see Figs. 8(b)(ii)-(v)]. In particular, for sufficiently large drift velocities, the symmetry breaking caused by the drift-current biasing gives rise to wide frequency bands wherein the acoustic and/or optical SPP modes only propagate along the $+x$ direction (the direction of the drifting electrons), consistent with the results of Sect. III for the drift-current biased SLG system.

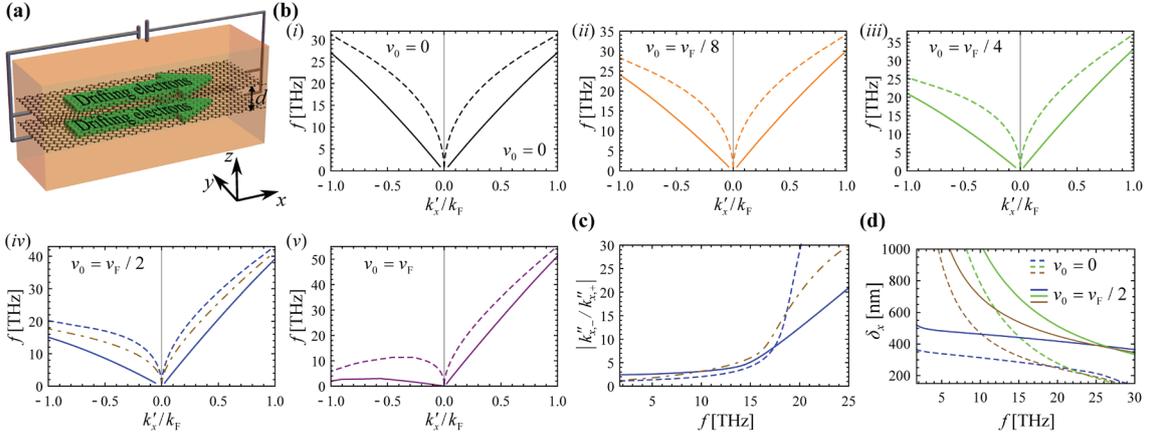

**Fig. 8.** (a) Two graphene sheets are embedded in a dielectric with permittivity $\varepsilon_{r,s}$. The graphene sheets are biased with identical drift currents. (b) (i)-(v) Dispersion of the SPPs supported by the DLG system for $d = 5$ nm and for the values of $v_0$ indicated in the insets. (c) Ratio between the attenuation constants of the plasmons that propagate along the $-x$ and $+x$ directions as a function of the frequency for $v_0 = v_F/2$. (b)-(c) Solid curves: acoustic (even) modes; dashed curves: optical (odd) modes; The brown dot-dashed curves in (b-iv) and (c) correspond to the SPP supported by the SLG configuration. (d) SPP propagation lengths ($\delta_x$) with (solid curves) and without (dashed curves) drift-current biasing as a function of the frequency for the SPPs propagating along the $+x$ direction. Blue curves: acoustic (even) mode; green curves: optical (odd) mode; brown curves: SPP SLG.

Moreover, analogous to what happens in the drift-current biased SLG system, the attenuation factor of the current-driven acoustic and optical SPPs that propagate along the $-x$ direction is considerably larger than that of the SPPs propagating along the $+x$ direction [see Fig. 8(c) for $v_0 = v_F/2$]. Hence, the unidirectional propagation of the DLG plasmons may be also achievable for drift velocities significantly smaller than $v_F$.

To further study how the drift-current biasing affects the dispersion properties of the DLG plasmons, we calculated the SPP propagation length [see Fig. 8(d)]. Clearly, the propagation length of both the acoustic and optical SPPs is considerably augmented when a drift-current biasing is applied to the graphene sheets.

The brown curves in Figs. 8(b-*iv*), (c) and (d) are for the SPP supported by a SLG configuration with $v_0 = v_F/2$. Quite interestingly, for frequencies above 19 THz $\left| k''_{x,-} / k''_{x,+} \right|$ has a maximum value for the optical SPP of the DLG system. This indicates that with two sheets of drifting electrons it may be possible to have an enhanced drag effect and enhanced optical isolation, as compared to a single graphene sheet. Furthermore, as seen in Fig. 8(d) the propagation length of the optical SPP of the DLG configuration is greater than for the SLG configuration up to frequencies on the order of 25 THz. These results suggest that the additional drift-current biased graphene sheet can enable a stronger nonreciprocity and better optical isolation, if the DLG system can be operated using the optical SPP.

## V. CONCLUSIONS

In conclusion, we theoretically studied the impact of the spatial dispersion of the bare graphene conductivity on the propagation of current-driven graphene plasmons in single-layer and double-layer graphene systems. It was shown that even though on the overall the nonlocality weakens the nonreciprocal effects, the drift-current bias remains a rather exciting solution to obtain one-way propagation and a one-atom thick "isolator". Interestingly, we have shown that the nonreciprocity strength and isolation level can be higher in a system formed by two drift-current biased graphene sheets, due to the additional drag provided by the drifting electrons of the second graphene layer. Furthermore, it was demonstrated that the drift-current biasing may strongly suppress the backscattering caused by an obstacle placed along the propagation path of the graphene SPPs and may boost the propagation length of the graphene plasmons. The drift-current biased graphene may be, therefore, a very attractive tunable nonreciprocal platform for highly-integrated nanophotonic circuits.

## ACKNOWLEDGMENTS

This work was partially funded by the IET under the A F Harvey Prize 2018, by Fundação para a Ciência e a Tecnologia (FCT) under the project PTDC/EEI-TEL/4543/2014, and by Instituto de Telecomunicações (IT) under project UID/EEA/50008/2019. T. A. Morgado acknowledges financial support by FCT under the the CEEC Individual 2017 contract as assistant researcher with reference CT/Nº004/2019-F00069 established with IT – Coimbra.

## APPENDIX A: ANALYTICAL CONTINUATION OF THE NONLOCAL CONDUCTIVITY FORMULA

In this Appendix, we extend the nonlocal zero-temperature conductivity formula (1) reported in [34-35] to the cases where either $k_x$ or $\omega$ are complex.

For a complex-valued $k_x$ ($k_x = k_x' + ik_x''$) with $|k_x''/k_x'| \ll 1$, $\sigma_g$ can be calculated using

$$\sigma_g(\omega, q) = -i\omega \frac{e^2}{4\pi\hbar} \left[ \frac{8k_F}{q^2 v_F} + \frac{G_{m,2}(\Delta_-) - (G_{m,2}(\Delta_+) - i\pi)}{S_{UHP}(\omega - qv_F) S_{UHP}(\omega - qv_F)} \right], \quad \text{(A1a)}$$

$$S_{UHP}(x) = \begin{cases} -\sqrt{x}, & \text{Re}[x] < 0 \text{ and } \text{Im}[x] < 0 \\ \sqrt{x}, & \text{other cases} \end{cases}, \quad \text{(A1b)}$$

$$G_{m,2}(z) = z\, S_{UHP}(z-1) S_{UHP}(z+1) - \ln\left( \left(z + S_{UHP}(z-1) S_{UHP}(z+1)\right) e^{i\pi/4} \right) - i\pi/4, \quad \text{(A1c)}$$

In the above, the square root and logarithm functions are the standard ones with branch cuts in the negative real axis. It can be checked that the formula provides an analytical continuation of Eq. (1), when $|k_x''/k_x'| \ll 1$.

Furthermore, when $k_x$ is real-valued Eq. (1) can be extended analytically to the first quadrant of the complex plane $\text{Re}\{\omega\} > 0$ and $\text{Im}\{\omega\} \geq 0$ as:

$$\sigma_g(\omega, q) = -i\omega \frac{e^2}{4\pi\hbar} \left[ \frac{8k_F}{q^2 v_F} + \frac{G(\Delta_-) - (G(\Delta_+) - i\pi)}{\sqrt{\omega^2 - q^2 v_F^2}} \right], \quad (A2)$$

with $G(z) = z\sqrt{z-1}\sqrt{z+1} - \ln(z + \sqrt{z-1}\sqrt{z+1})$. The square root and the logarithmic functions are again the standard ones, with branch cuts in the negative real-axis (note that due to this reason, in general $\sqrt{z-1}\sqrt{z+1} \neq \sqrt{z^2 - 1}$). Furthermore, Eq. (A2) can be extended to the entire upper-half frequency plane using the reality condition:

$$\sigma_g(\omega, q) = \sigma_g^*(-\omega^*, q^*). \quad (A3)$$

**APPENDIX B: ENERGY STORED IN GRAPHENE**

The instantaneous power transferred to the electrons in the graphene by the electromagnetic field is given by $p = \int ds\, \mathbf{j}_s \cdot \mathbf{E}$. Here, $\mathbf{E}$ is the instantaneous electric field and $\mathbf{j}_s$ the surface electric current density in graphene. The integral is over the area of the graphene sheet. For a time harmonic excitation with frequency $\omega = \omega' + i\omega''$, the relevant fields are of the form $\mathbf{E} = \frac{1}{2}\left(\mathbf{E}_\omega e^{-i\omega t} + \mathbf{E}_\omega^* e^{+i\omega^* t}\right)$ and $\mathbf{j}_s = \frac{1}{2}\left(\sigma_g(\omega)\mathbf{E}_\omega e^{-i\omega t} + \sigma_g(-\omega^*)\mathbf{E}_\omega^* e^{+i\omega^* t}\right)$. Thus, the transferred power is:

$$\begin{aligned} p &= \frac{1}{4}\int ds \left(\mathbf{E}_\omega e^{-i\omega t} + \mathbf{E}_\omega^* e^{+i\omega^* t}\right) \cdot \left(\sigma_g(\omega)\mathbf{E}_\omega e^{-i\omega t} + \sigma_g(-\omega^*)\mathbf{E}_\omega^* e^{+i\omega^* t}\right) \\ &= \frac{1}{4}\int ds \left(|\mathbf{E}_\omega|^2 \left[\sigma_g(\omega) + \sigma_g(-\omega^*)\right] e^{+2\omega'' t} + ...\right) \end{aligned} \quad (B1)$$

where the term "…" has a time variation of the type $e^{2\omega'' t} e^{\pm i 2\omega' t}$. For a lossless system, $p = \frac{d\mathcal{E}}{dt}$ with $\mathcal{E}$ the "stored" (kinetic) energy. Thereby, the change of the electrons kinetic energy due to the electromagnetic interaction is:

$$\delta \mathcal{E} = \frac{1}{4} \int ds \left( |\mathbf{E}_\omega|^2 \left[ \sigma_g(\omega) + \sigma_g(-\omega^*) \right] \frac{1}{2\omega''} e^{+2\omega''t} + ... \right), \tag{B2}$$

where the term "..." has a time variation of the type $e^{2\omega''t} e^{\pm i 2\omega't}$. Hence, letting $\omega'' \to 0$ it follows that the time-averaged "stored" energy in time-harmonic regime is given by:

$$\begin{aligned}
\delta \mathcal{E}_{av} &= \lim_{\omega'' \to 0} \frac{1}{4} \int ds \, |\mathbf{E}_\omega|^2 \left[ \sigma_g(\omega' + i\omega'') + \sigma_g(-\omega' + i\omega'') \right] \frac{1}{2\omega''} \\
&= \lim_{\omega'' \to 0} \frac{1}{4} \int ds \, |\mathbf{E}_\omega|^2 \left[ \sigma_g(\omega') + \sigma_g(-\omega') + i 2\omega'' \partial_\omega \sigma_g(\omega') \right] \frac{1}{2\omega''}, \\
&= \frac{1}{4} \int ds \, |\mathbf{E}_\omega|^2 \mathrm{Im}\{-\partial_\omega \sigma_g(\omega')\}
\end{aligned} \tag{B3}$$

We used the properties $\sigma_g(\omega') + \sigma_g(-\omega') = 0$ and $\sigma_g(\omega') = i \mathrm{Im}\{\sigma_g(\omega')\}$ for a lossless graphene (i.e., below the interband transition threshold). Therefore, for a lossless passive graphene it is necessary that $\mathrm{Im}\{\partial_\omega \sigma_g(\omega')\} < 0$ so that the energy transferred from the field to the electrons is positive. When the conductivity dispersion is anomalous, $\mathrm{Im}\{\partial_\omega \sigma_g(\omega')\} > 0$, the moving electrons give away their energy to the field, rather than extracting energy from it, until the steady state is reached. This can only occur in the presence of a drift current.

## APPENDIX C: RELATIVISTIC DOPPLER SHIFT MODEL OF THE GRAPHENE CONDUCTIVITY

The nonlocal polarizability of a drift-current biased graphene sheet was derived in Ref. [26] relying on a single band model and a skewed Fermi distribution. Based on such a model the authors derived a dispersion equation for the graphene plasmons [Eq. (1), 26]. The corresponding formula for the graphene conductivity is not written explicitly in Ref. [26] but can be found by comparing the quasi-static dispersion of the graphene plasmons ($|k_x| - 2\varepsilon_{r,s} \kappa_g^{\mathrm{drift}} = 0$ with $\kappa_g^{\mathrm{drift}}(\omega, k_x) = i\omega\varepsilon_0 / \sigma_g^{\mathrm{intra,drift}}(\omega, k_x)$) with Eq. (1) of Ref. [26]. This yields the result:

$$\sigma_{\text{g}}^{\text{intra,drift}}(\omega, k_x) = \frac{i\omega e^2}{\hbar^2} \frac{2\mu_c}{\pi} \frac{1}{\gamma(\omega - k_x v_0)\sqrt{\omega - k_x v_F}\sqrt{\omega + k_x v_F} + \omega^2 - v_F^2 k_x^2}, \quad (C1)$$

with $\gamma = 1/\sqrt{1 - v_0^2/v_F^2}$ the graphene Lorentz factor. In the absence of a drift-current, the conductivity reduces to:

$$\sigma_{\text{g}}^{\text{intra}}(\omega, k_x) = \frac{i\omega e^2}{\hbar^2} \frac{2\mu_c}{\pi} \frac{1}{\omega\sqrt{\omega - k_x v_F}\sqrt{\omega + k_x v_F} + \omega^2 - v_F^2 k_x^2}. \quad (C2)$$

Interestingly, the $\sigma_{\text{g}}^{\text{intra,drift}}(\omega, k_x)$ is related to $\sigma_{\text{g}}^{\text{intra}}(\omega, k_x)$ through a relativistic Doppler shift transformation [Eq. (5)] such that $\sigma_{\text{g}}^{\text{intra,drift}}(\omega, k_x) = (\omega/\tilde{\omega})\sigma_{\text{g}}^{\text{intra}}(\tilde{\omega}, \tilde{k}_x)$, with $\tilde{\omega} = \gamma(\omega - k_x v_0)$ and $\tilde{k}_x = \gamma(k_x - \omega v_0/v_F^2)$. The no-drift nonlocal intraband conductivity of graphene [Eq. (C2)] can also be obtained using Boltzmann's theory [37]. In the $k_x \to 0$ limit, Eq. (C2) reduces to the standard Drude model of graphene $\sigma_{\text{g}}^{\text{intra}}(\omega, k_x \to 0^+) = \frac{ie^2}{\hbar^2}\frac{\mu_c}{\pi}\frac{1}{\omega}$ [41].